\documentclass[]{spie}  %>>> use for US letter paper
\usepackage{amsmath,amsfonts,amssymb}
\usepackage{graphicx}
\usepackage[colorlinks=true, allcolors=blue]{hyperref}

\title{WUVS Simulator: Detectability of spectral lines with the WSO--UV spectrographs}

\author[a]{Pablo Marcos-Arenal}
\author[a]{Ana I. G\'omez de Castro}
\author[a]{Bel\'en Perea Abarca}
\author[b]{Mikhail Sachkov}
\affil[a]{AEGORA Research Group, Universidad Complutense de Madrid, Spain}
\affil[b]{Institute of Astronomy RAS, Moscow, Russian Federation}

\authorinfo{Further author information: (Send correspondence to P.M.A.)\\P.M.A.: E-mail: pablmarc@ucm.es, +34 91 394 40 62}

\begin{document} 
\maketitle

\begin{abstract}
The World Space Observatory  -- Ultraviolet (WSO--UV) space telescope is equipped with high dispersion (55,000) spectrographs working in the 1150-3100 \AA ~spectral range.
To evaluate the impact of the design on the scientific objectives of the mission, a simulation software tool  has been developed. 
This simulator builds on the development made for the PLATO space mission and it is designed to generate synthetic time-series of images by including models of all important noise sources.
In this article, we describe its design and performance. 
Moreover, its application to the detectability of important spectral features for star formation and exoplanetary research is addressed.
 
\end{abstract}


\section{INTRODUCTION}
\label{sec:intro}  % \label{} allows reference to this section

The World Space Observatory -- Ultraviolet (WSO--UV) is a 170 cm aperture space telescope to observe in the ultraviolet (UV) range\cite{Sachkov2014}. 
WSO--UV is equipped with instrumentation for ultraviolet (1150-3100 \AA ) imaging and spectroscopy with an extension to the optical range (up to 6000 \AA ). 
WSO--UV is a Russian led mission that will be operating in high-Earth orbit (geosynchronous with inclination 51.6$^{\circ}$) for five years with a possible extension to  five more years\cite{Shustov2014}, providing access to the UV range to the world-wide astronomical community. 
Spain is a major partner to the mission providing  software and support for the science and mission operations.
The WSO--UV spacecraft is equipped with three spectrographs; all together are named WUVS (WSO--UV spectrographs) \cite{Panchuk2014}:

\begin{itemize}
	\item  The far UV high resolution spectrograph (VUVES) to carry out high resolution spectroscopy ($R \sim 55,000$) in the range 1150-1760 \AA .
	\item  The near UV high resolution spectrograph (UVES) for high resolution spectroscopy  ($R \sim 55,000$) in the range 1740-3100 \AA.
	\item  The Long Slit Spectrograph (LSS)  will provide low resolution ($R \sim 1,000$), long slit spectroscopy in the range 1150-3050 \AA. The spatial resolution will be 0.5 arcsec.
\end{itemize}

The entrance slits lie on a circle of diameter 100 mm on the focal plane, underneath the imagers (see Ref.\cite{Sachkov2014} for a description of WSO--UV focal plane).
WUVS  will be equipped with CCD detectors, providing high sensitivity, high geometrical stability and high dynamic range \cite{Shugarov}.
For cost-saving reasons the design of all three detectors is  identical except for minor changes like the anti-reflection coating on the CCD or the selection of active output amplifiers.

WSO--UV will run three main scientific programs: core, national and open. 
The national program will grant observing time to Russian and Spanish observers, and the open program will grant access to the world wide scientific community. 
Both will work on competitive basis selecting the best scientific projects proposed by each community. 
The core program is designed to run the key scientific projects for WSO--UV and will run only during the first three years of the mission.
This program needs to be developed well in advance and plays a key role in the definition of the scientific requirements for the instrumentation development.
Software tools are being developed by the consortium to ease the feedback loop between scientific investigators and instrument developers. 
The WUVS simulator (or WUVS--Sim) has been developed for this purpose. 
Based on the heritage of the simulator developed for the PLATO mission \cite{Marcos-Arenal2014b}, WUVS--Sim deals efficiently with noise sources handling and will be used in the calibration process definition. 

In this article we describe the development status of WUVS--Sim. 
In Section~2, the simulator is described; in Section~3 simulations of WUVS expected performance for the detectability of spectral features are shown.
A brief summary is provided at the end, in Section~4.

%#####################################################################################
\section{WUVS Simulator}
\label{sec:development}

WUVS--Sim has been implemented as a further development of the PLATO Simulator\cite{Marcos-Arenal2014b} (PLATOSim). 
PLATO is a medium-class space mission approved by the European Space Agency (ESA), aimed to find and study extrasolar planetary systems with emphasis on the properties of Earth-like planets in the habitable zone around solar-like stars {\bf \cite{Rauer2013}}.  
PLATO is equipped with 24 identical cameras (12 cm aperture) and 96 CCDs (four CCDs per camera) with $4510\times 4510$ $18 \mu $m pixels. PLATOSim is an end-to-end software tool developed at the Institute for Astronomy at KU Leuven for the validation of the noise levels requirements of the PLATO mission, but designed to be easily adaptable  to similar types of  missions\cite{Marcos-Arenal2014a}. 
PLATOSim is open source and available upon request.
It has been developed in C++  programing language to be used under Linux platforms.

PLATOSim generates time-series of realistic images of the foreseen observations by including models of the CCD and its electronics, the telescope optics, the stellar field, the jitter movements of the spacecraft and all important natural noise sources. 
Inputs to PLATOSim are the equatorial coordinates of the astronomical sources in the field of view, their magnitudes and the characteristics of the optics and  detection chain. 
PLATOSim output includes the photometric analysis of the sources in the generated images,  their magnitude, and an estimate of the noise level.  
PLATOSim was applied to validate the noise level requirements of the mission: $<34$~ parts-per-million (ppm) per hour for stars with visual magnitudes $< 11$th mag, and $<80$~ppm per hour for stars with visual magnitudes  $> 13$th mag.  

Several code modifications were implemented to PLATOSim in order to adapt it to WUVS, and more specifically to the VUVES instrument. 
WUVS--Sim allows evaluating noise level response, data quality, observing strategies, and instrument fine-tuning design for different types of configurations. 
It is designed to test the scientific feasibility of an observing proposal and develop the data processing pipeline before the actual spacecraft is placed into orbit and fully operational.

\subsection{Task flow in WUVS--Sim}

WUVS--Sim generates synthesized images by simulating the acquisition process of a space-based instrument as realistically as possible. 
Each image is numerically modeled from a number of input parameters which define the set-up of the CCD and its electronics, the properties of the optical instrument, the pixel response non-uniformity (PRNU) and all related noise sources.
The process of image generation follows as per this sequence: 

\begin{enumerate}
	\item Input image at the detector plane. 
	\item Image acquisition by the CCD detector: CCD sub-pixel breakdown;
	\item ICCD Sensitivity variations: PRNU;
	\item Noise effects:
%		\subitem Read out smearing; 
		\subitem Sky background;
		\subitem Photon noise; 
		\subitem Other electronic noise sources.
\end{enumerate}

%#####################################################################################

\subsection{Input parameters}

The main modification implemented to PLATOSim is to accept  FITS files as input, instead of a star catalogue. 
PLATOSim generates synthetic images of the Field of View (FoV) taking the right ascension, declination and magnitude of each source, and transferring their position to the image FoV; afterwards, it calculates the number of counts (also called Analog-to-Digital Units, or $ADU$) to be registered by each pixel on the detector, depending on the magnitude of the source.  
This modification decouples the optical segment from the detection segment in the Simulator, to make the computational scheme more efficient for an echelle spectrograph as WUVS.
As a side effect, the impact of jittering or the charge transfer smearing are not included in the current version of WUVS--Sim since they were determined in PLATOSim at the time of 
calculating the position (pixel)  of each source.

All other noise effects included in our simulations were modeled by using the parameters described in Table \ref{tab:input_params}.
For the simulations in this work,  thermal noise has been considered negligible.

%__________________TABLE__________________________________________________________
\begin{table}[ht]
\caption{Values of the input parameters applied to the WUVS simulations.} 
\label{tab:input_params}
\begin{center}       
\begin{tabular}{|l|c|} 

 \hline
\emph{Input Parameter} & \emph{Value}  \\
 \hline
CCD Size & $4096 \times 3112 \, px$ \\
%Collecting area &113.09 $cm^{2}$\\
%Sub-Field size & $4096 \timesÃÂ 3112 \, px$ \\
Digital saturation    & 16384 $ADU$   \\ 
Pixel resolution   & 1/4 \\
Full well pixel capacity   &1243000 $\mathrm{e^-}$ \\
%Transmission efficiency &0.638466 \\
Gain  & 6  $\mathrm{e^-}/ADU$ \\ 
%Quantum efficiency   & 85 ${\%}$ \\
Electronic offset   &100 $ADU$ \\
%Number of exposures   &1000  \\
Readout noise   &  3 $\mathrm{e^-}$ \\
Exposure time &22 $s$\\
Flatfield pixel-to-pixel noise  & 1.6 ${\%}$  \\ 
%Charge transfer time    & 3 $s$ \\
%Mean Charge Transfer Efficiency   &0.999999 \\
Pixel scale  & 0.101 $arcsec/px \hspace{0.02\linewidth}$ \\
 Pixel size   &  12 $\mu m$  \\ 
\hline
\end{tabular}
\end{center}
\end{table} 
%_______________________________________________________________________________

\subsection{WUVS--Sim validation}

\begin{figure} [ht]
   \begin{center}
   \includegraphics[width=\linewidth, trim=0 0 0 0, clip]{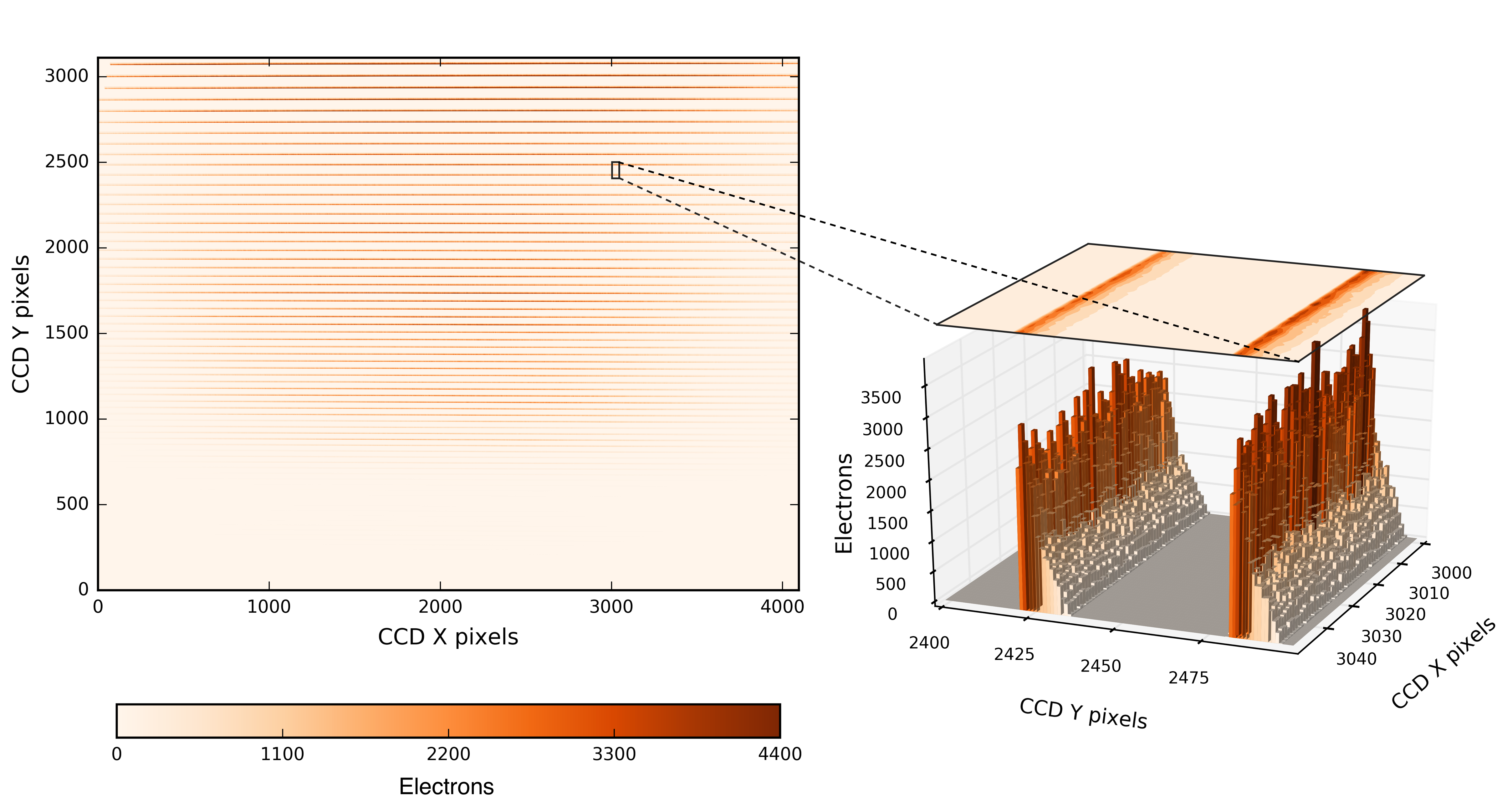}
   \end{center}
   \caption[] 
   {\label{fig:fig1}(Left) Input image to WUVS--Sim: VUVES dispersal of the radiation from a lamp with a flat energy distribution.
    The echelle  pattern is visible as the horizontal stripes for each spectral order. 
     (Right) 3D enlarged image of a small piece of 10th and 11th orders.  }
   \end{figure}

We performed a set of tests in order to evaluate the behavior of the electronic and photonic noise in the simulated images.  
For this purpose, we have used as input  the image produced by WUVS when illuminated by a lamp with a flat energy distribution in VUVES spectral range (see Fig.~\ref{fig:fig1}). 
This $4096 \times 3112 $ pixels  image (hereafter, flat image) have been computed by the WUVS instrument team to study of the optical performance of the instrument  (see Ref.~\cite{Panchuk2014} for a detailed description of the optical layout of the WUVS spectrographs). 
The optical pattern produced by an echelle spectrograph is clearly recognizable as well as the ripples at the edges of the orders. 
The spectral order number decreases as the Y coordinate in Fig.~1 increases. 

%_______fig1________________________________________________________________________
%_______________________________________________________________________________

Fig.~\ref{fig:fig1} (right panel) displays in 3D a small area of orders 10 to 11 more precisely, from X=3000 to X=3050  pixels and from Y=2400 to Y=2500.   
The Z axis represents the number of electrons produced by the photons impinging on the CCD surface; there is a noticeable interorders tail produced by the optics. To guarantee photometric accuracy, the tail must be included in the flux calculation.
However, it adds noise to the final results as shown in the figure. 

\begin{figure} [ht]
   \begin{center}
   \includegraphics[width=\linewidth, trim=70 0 60 35, clip]{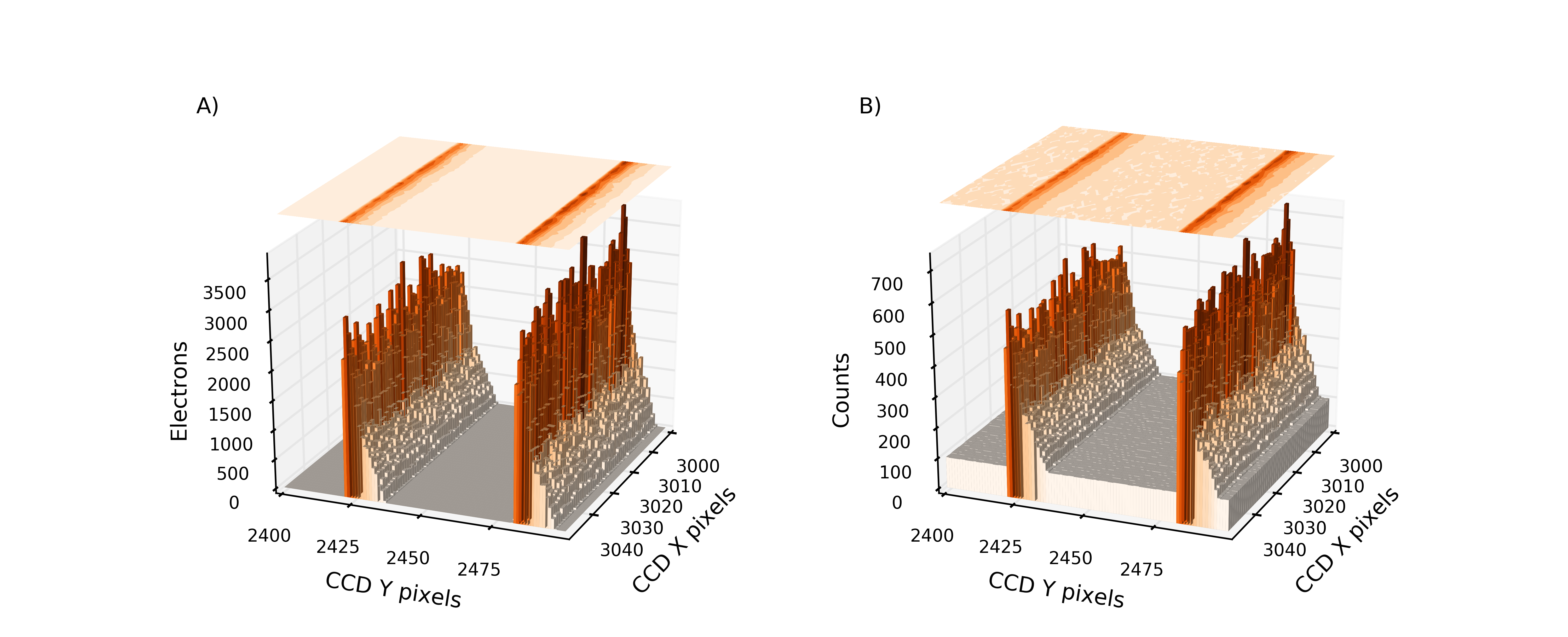}
   \end{center}
   \caption[] 
   {\label{fig:fig2}  3D enlarged flat  image of a small piece of 10th and 11th orders.  
Panel A: Flat input image.  Panel B: Flat simulated image as output of the WUVS--Sim. }
   \end{figure}

As a test, several simulations have been run with WUVS--Sim and all of them reproduce the same images unless for the photonic white noise; one sample is shown in Fig.~\ref{fig:fig2}. 
Left panel in the figure shows the flat input image, while right panel shows the flat simulated image as output of the WUVS--Sim.
The impact of the detection electronics is  identifiable in the electronic offset (set to 100 ADU,  see Table \ref{tab:input_params}) in the background of the flat simulated image. 
It can also be noted the gain value, set to 6  $\mathrm{e^-}/ADU$ (see Table \ref{tab:input_params}). 

%_______fig2_______________________________________________________________________

%_______________________________________________________________________________

These images are not showing any saturation effect since the input image is far from saturation, but WUVS--Sim will be used to study saturation and charge bleeding impact in the nearby continuum, spectral lines and interorder background calculation. 
This is also applicable to the high energy particle hits. 
It will be required a dedicated set of simulations and analysis of their outcome to see how these effects would affect the observations. 
Anyhow, it must be noted that the inclusion of any of these effects will degrade these simulated images and the results presented in chapter  \ref{sec:outcome}.
This analysis is required to optimize the pipeline processing of WSO--UV data.

%#####################################################################################
\section{Simulation of emission lines spectra with WUVS--Sim: the low intensity limit}
\label{sec:simulations}

We performed an additional set of simulations to evaluate the impact of the CCD performance on the detectability of weak spectral lines. 

\subsection{Source selection and input spectrum}
\label{sec:inputspectrum}

In order to analyze the simulator performance, we modified the echelle image of a flat illuminated source (Fig.~\ref{fig:fig1}) to include the real  observation obtained with the Space Telescope Imaging Spectrograph (STIS)\cite{Biretta2016} of the T Tauri star DG~Tau.
DG~Tau far UV spectrum is dominated by emission lines;   Lyman-$\alpha$ (1216 \AA), O~I(1306 \AA), C~II(1335 \AA), Si~IV(1400 \AA), C~IV (1550 \AA) and C~I(1670 \AA)  are prominent resonance transitions (see Ref. \cite{Ardila2013} or Ref. \cite{Gomez2013}). 
These lines are of interest for star formation and exoplanetary research.

DG~Tau has been observed with STIS, using the E140M grating that provides echelle spectra in the range 1144-1710 \AA~ with resolving power  $ R \sim 45800$, similar to VUVES. 
The spectrum is noisy and well suited for our numerical test.

The DG Tau STIS spectrum was normalized in flux and corrected in terms of spectral resolution per pixel according to the VUVES resolution (0.0066 \AA/px). 
Then, we superimposed the spectral orders of interest from STIS (121st for Lyman-$\alpha$, 113th for O~I, 111th for C~II and  89th for C~I) on the corresponding pixel positions of the flat echelle image of VUVES (7th for Lyman-$\alpha$, 10th for O~I, 11th for C~II and  24th for C~I). 
%It is not needed to include all others STIS spectral orders for our study.    

\begin{figure} [ht]
   \begin{center}
   \includegraphics[width=\linewidth, trim=70 30 60 0, clip]{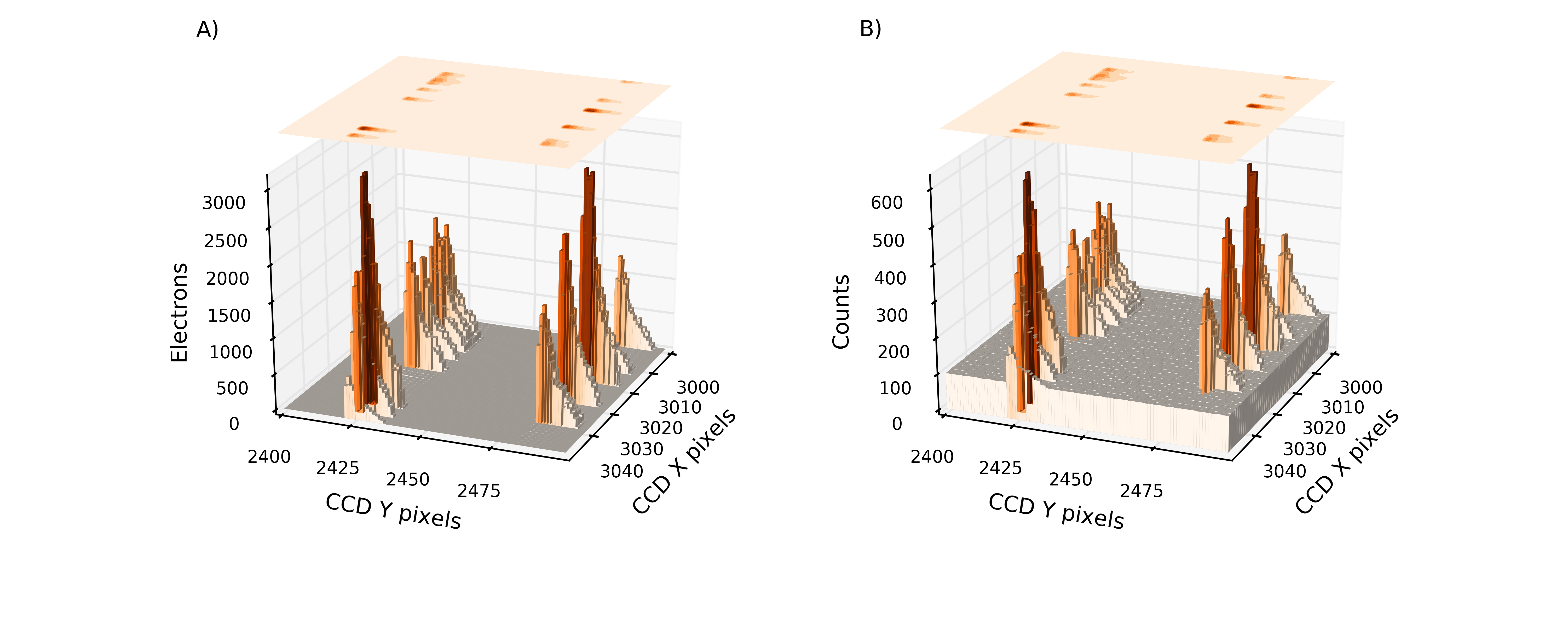}  
   \end{center}
   \caption[]
   { \label{fig:fig3} 
     3D enlarged echelle image of a small piece of DG Tau of 10th and 11th orders at pixel level:
     Panel A:  DG Tau echelle input image. Panel B:  DG Tau echelle simulated image as output of the WUVS--Sim.  }
   \end{figure}

The resulting image is taken as input to the VUVES simulations (see Fig.~\ref{fig:fig3}-A, showing a fraction of this input image). 
Fig.~\ref{fig:fig3}-B is a fraction the output of the simulation, and shows the effect of electronic offset (set to 100 ADU) and the gain value (electrons to ADU set to 6  $\mathrm{e^-}/ADU$). 
As expected from the input parameters shown in Table \ref{tab:input_params}.

%_______fig3_______________________________________________________________________
%_______________________________________________________________________________

\subsection{WUVS--Sim based analysis of selected spectral lines}
\label{sec:outcome}

From WUVS--Sim output images (see Sect.~\ref{sec:inputspectrum}), we selected some few features (Lyman-$\alpha$, O~I, C~I and C~II)  to analyze WUVS spectral response.
These lines are of interest because they cover a wide range of strengths and wavelengths ({\it i.e.} locations in the CCD detector). In particular, Lyman-$\alpha$ is in order 7th in WUVS, O~I in 10th, C~II in 11th and C~I in 24th.   

\begin{figure} [ht]
   \begin{center}
   \includegraphics[width=\linewidth, trim=80 25 65 40, clip]{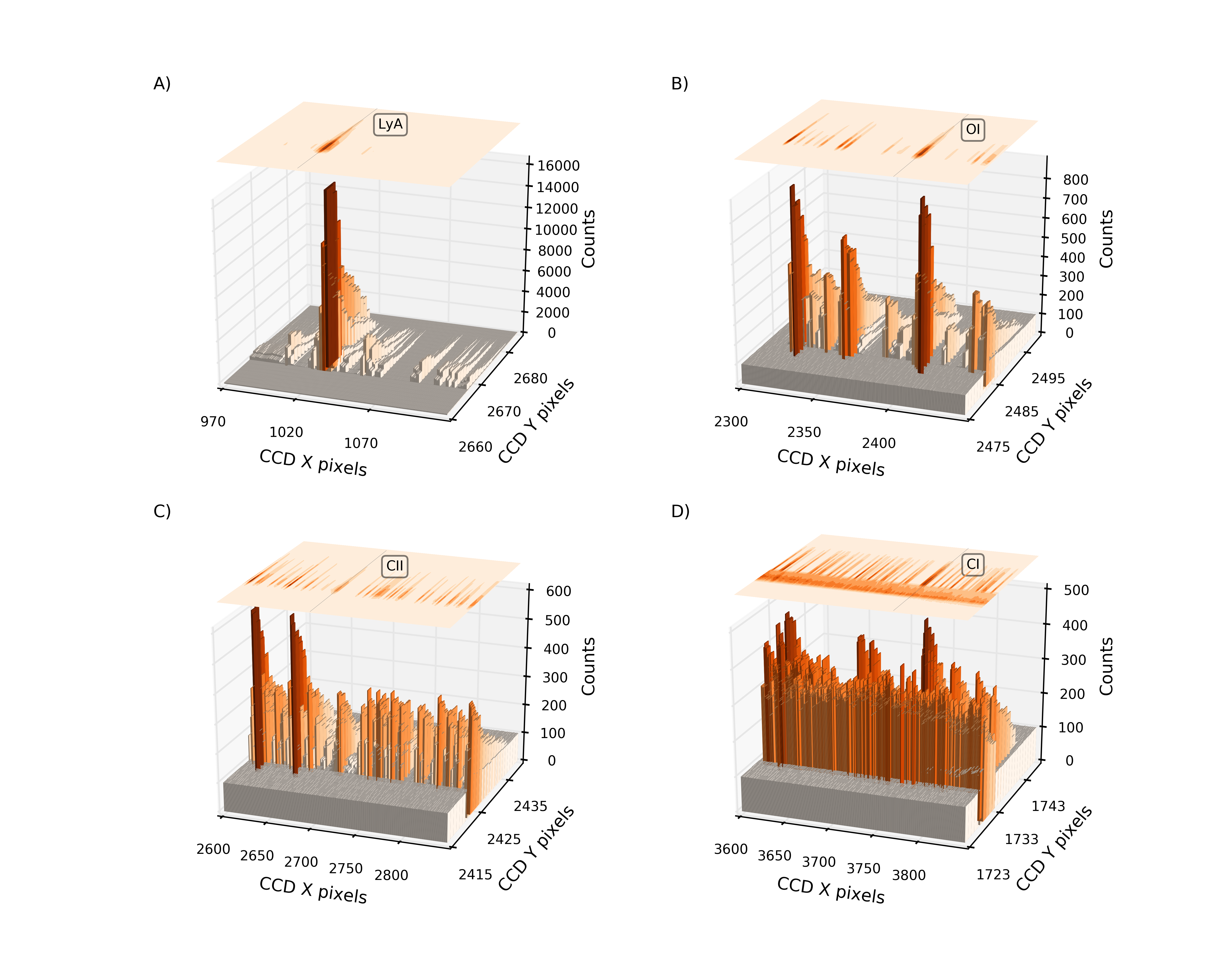}  
   \end{center}
   \caption[]{ \label{fig:fig4}
   3D fractions of a DG Tau echelle simulated image for the spectral orders corresponding to Lyman-$\alpha$ (LyA, panel A), O~I (panel B), C~II (panel C) and C~I (panel D).} 
   \end{figure}

Fig.~\ref{fig:fig4} shows the enlargement of WUVS--Sim echelle simulated images for the spectral orders corresponding to Lyman-$\alpha$ (LyA, panel A), O~I (panel B), C~II (panel C) and C~I (panel D). 
The number of counts for Lyman-$\alpha$ (Fig.~\ref{fig:fig4}-A) reaches the digital saturation level (16384 $ADU$; see Table \ref{tab:input_params}) at some pixels, running over to the adjacent pixels upwards in the CCD charge transfer direction (Y axis). 
O~I and C~II (Fig.~\ref{fig:fig4}-B and Fig.~\ref{fig:fig4}-C, respectively) are weaker features, but their spectral lines are still noticeable in this raw pixel representation. 
Finally, we chose C~I for two reasons: Firstly, its flux is quite low in the original STIS spectrum, and secondly, it is located on the extreme of the 24th spectral order of VUVES (around pixel 3750 in the X axis). 
The detected flux for C~I (Fig.~\ref{fig:fig4}-D) is enough for identifying its spectral line, but the highlight on this plot is the continuum-like stripe in the lower end of the spectrum.  
This is due to the decrease in optical performance of the spectrographs at the extremes of each spectral order. 
Fig. \ref{fig:fig1}  shows the throughput decrease at the extremes, and at higher spectral orders (lower values in the CCD Y axis).
But there is also a decrease in the optical quality at these ranges, as it is shown in Fig.~\ref{fig:fig4}-D.

%_______fig4________________________________________________________________________
%_______________________________________________________________________________

For the sake of the numerical experiment, we applied only  basic data reduction process to the output echelle simulated images (hereafter referred to as \textit{simulated spectra});
no wavelength calibration has been applied since  we could apply the same wavelength solution that we applied to set the  STIS spectral orders of DG Tau in the  input echelle image.
Further development of WUVS--Sim into WUVS pipeline tester will be implemented as the WSO-UV develops.

\begin{figure} [ht]
    \begin{center}
   \includegraphics[width=\linewidth, trim=90 0 100 0, clip]{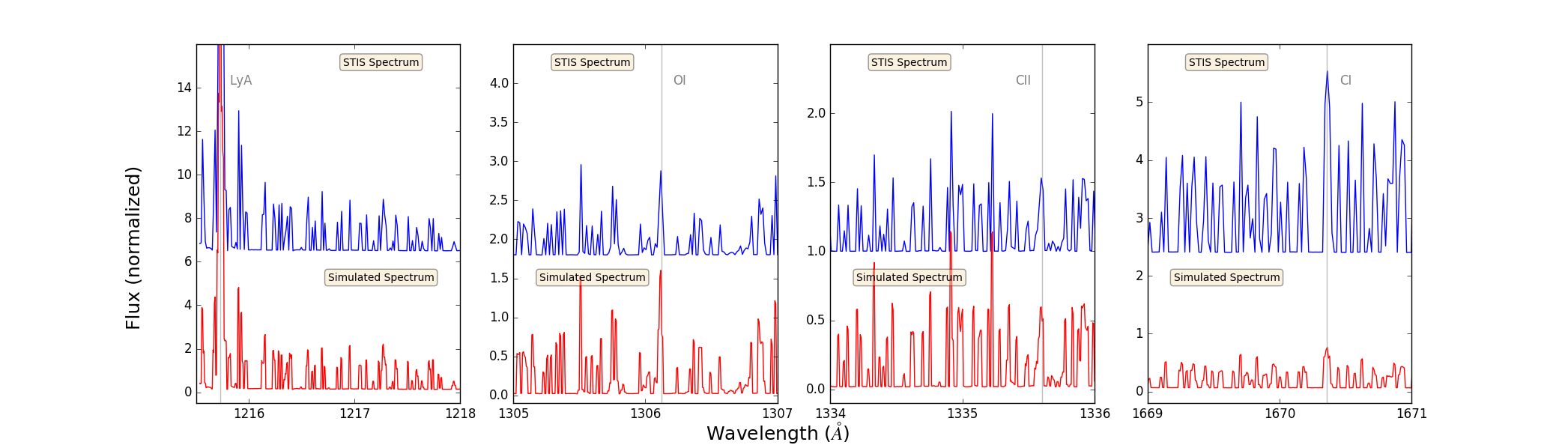}  
    \end{center}
    \caption[]{ \label{fig:reduced_spectra} 
    Normalized flux as a function of wavelength for four spectral lines (Lyman-$\alpha$, C~I, C~II and O~I) of DG Tau as obtained from STIS (in blue color) and from the echelle simulated  image (in red color). 
    The STIS spectra have been shifted upwards for comparison purposes.} 
    \end{figure}

The final measured flux for each wavelength is obtained by adding the values (counts) of the pixels in the same column (perpendicular  to the dispersion direction).
This is displayed in Fig. \ref{fig:reduced_spectra}  for Lyman-$\alpha$, O~I, C~II and C~I (in red color) and compared with STIS source spectra (blue color).
The flux in each of the represented spectra has been normalized to the mean flux of its corresponding spectral order to ease the comparison.

Fig. \ref{fig:reduced_spectra} shows small differences between the original and the \textit{simulated spectra} in Lyman-$\alpha$, O~I, and C~II spectral orders, but a highly noticeable decrease in flux  in C~I caused by the decreasing efficiency of the optical system in the edge of the echelle orders  (see Fig. \ref{fig:fig1}). 
But it is also noticeable that there is an small broadening in the C~I peak, due to the decrease in the optical quality at these ranges, as it is shown in Fig.~\ref{fig:fig4}-D.

%_______fig5________________________________________________________________________
%_______________________________________________________________________________

The study presented here can easily be extended to other spectral lines. We are presenting these results as a use case of the simulator.
Fig. \ref{fig:reduced_spectra} shows  the power of WUVS--Sim to define and optimize WSO-UV programs of observation through a realistic  evaluation of detectability and  expected signal-to-noise or the required exposure time.

It is worth noting that the simulation output images require a certain data processing to obtain their spectra, meaning that the analysis will always be dependent upon the processing algorithm. 
Here we are applying a quite simple data reduction process, but this process can be improved in order to enhance the detected lines, but  this issue is out of the scope of this work.    

Finally, we would like to note that the main reason for the small difference between simulated and original spectrum is the absence of the jittering in the simulations. 
Simulations performed with PLATOSim proved jitter to be a main contributor to the overall noise budget \cite{Marcos-Arenal2014b}. 
Hence, next step in the development of the WUVS Simulator will be the implementation of the jitter contribution to the noise.  

%#####################################################################################
\section{CONCLUSIONS AND FUTURE PROSPECTS}
\label{sec:Conclusions}
WUVS instruments are high-precision spectrographs whose expected performance must be evaluated carefully from an appropriate overall instrument model. Since it is not feasible to build and test a prototype of a space-based instrument, numerical simulations performed by an end-to-end simulator are used to model the noise level expected to be present in the observations. 
The performance of the instrument can be evaluated in terms of noise source response, data quality, and fine-tuning of the instrument design for different types of configurations and observing strategies. In this way, a complete validation of the expected instrumental behavior of the mission can be derived.

As expected, the detectability of spectral lines would slightly depend upon the instrumental noise. 
Only those weak lines in the detectability limit would be affected by the instrumental noise. 
And even the detectability of these weak lines  would be improved by applying carefully designed observation strategies and data reduction processes. 
That is one of the aims of this work: providing  with a tool that can easily validate observations and optimize the observing strategies.      

In the future to come we are planning to apply the simulator to different astronomical sources in a wide range of brightness and spectral types. 
It will also be applied to evaluate the performance on the full wavelength range.  

It has already been stated how important is satellite jitter in the overall instrumental noise budget. 
We are working to implement this effect in the new release of the WUVS--Sim.

Many other improvements will be implemented to the WUVS Simulator in order to upgrade its usability and performance:
Generating an echelle spectra from standard libraries of high dispersion stellar spectra, thermal noise,  graphical user interface, and code parallelization among others.

% References
\bibliography{Marcos2016} % bibliography data in report.bib
%\bibliography{report} % bibliography data in report.bib
\bibliographystyle{spiebib} % makes bibtex use spiebib.bst

\end{document}